\documentclass[%
superscriptaddress, aps, pre
 amsmath,amssymb,
 aps, physrev,
]{revtex4-2}

\usepackage{floatrow}
\usepackage{etoolbox}
\usepackage{ulem}
\usepackage{xr}
\makeatletter
\patchcmd\captionlabel{\caption@setsubtype*{\FR@tmp}}{\caption@setsubtype*}{}{}
\makeatother

\newfloatcommand{capsidebox}{figure}[\capbeside\useFCwidth][\FBwidth]

\DeclareCaptionLabelFormat{rparen}{\noindent\llap{(\emph{#2})\,}}
\usepackage{adjustbox}

\makeatletter
\newcommand\footnoteref[1]{\protected@xdef\@thefnmark{\ref{#1}}\@footnotemark}
\makeatother

\usepackage{babel} 
\usepackage{csquotes}

\usepackage{amsthm,amsmath,amssymb}
\usepackage{mathrsfs}

\usepackage{algorithm}
\usepackage{algpseudocode}
\usepackage{amsmath}

\usepackage{amsthm}

\usepackage{mathtools}

\usepackage {graphicx}
\usepackage{caption}
\usepackage{epsfig}
\usepackage{epstopdf}
\usepackage{subcaption}
\usepackage{amsmath,bm}
\usepackage{color}
\usepackage{lmodern}
\usepackage[T1]{fontenc}
\usepackage{geometry}
\usepackage{subcaption}
\usepackage{bm}%
\usepackage[colorlinks=true,urlcolor=blue,citecolor=blue,linkcolor=blue]{hyperref}%
\usepackage{tabularx} 
\usepackage{booktabs}

\usepackage{multirow}

\usepackage{threeparttable}

\usepackage{everypage}

\usepackage{makecell}

\AddEverypageHook{\setcounter{footnote}{0}}

\expandafter\ifx\csname package@font\endcsname\relax\else
 \expandafter\expandafter
 \expandafter\usepackage
 \expandafter\expandafter
 \expandafter{\csname package@font\endcsname}%
\fi
\usepackage{verbatim}

\usepackage{xcolor}

\usepackage{float}
\usepackage{subcaption}
\usepackage[percent]{overpic}

\usepackage{booktabs}
\usepackage{siunitx}

\begin{document}


\title{\textbf{Inferring cell property distributions from high-throughput measurement data with cell-to-cell variability}}

\author{Stephan Gekle}
\email{stephan.gekle@uni-bayreuth.de}
\affiliation{Theoretische Physik VI, Universität Bayreuth, Bayreuth, Germany}

\begin{abstract}
High-throughput measurements of cell populations produce broad distributions of observables due to inherent cell-to-cell variability. 
While these distributions contain valuable biological information, extracting the underlying distributions of cellular properties remains challenging. 
We introduce the likelihood functional, a statistical framework that infers cell property distributions by combining experimental data with deterministic computational models. 
The method first evaluates the model on a discrete parameter grid. 
Optimization of the likelihood functional is then possible without repeated model evaluations and yields best-fit property distributions which enable the direct prediction of observable distributions for comparison with experiments. 
As a demonstration, we apply the framework to real-time deformability cytometry data and infer cytosol and membrane viscosity distributions of human red blood cells.

\end{abstract}

%

\maketitle

Biologists have long been aware that properties of individual cells can vary significantly across a population even of nominally identical cells.
As a direct consequence, high-throughput measurements on cell populations provide wide distributions of data rather than single values. 
These measured distributions contain important biophysical information as they directly reflect cell-to-cell variability within the population.
A number of theoretical approaches have been developed to infer cell property distributions from measured data based on Monte-Carlo sampling procedures \cite{Hasenauer2011, Loos2018, Dixit2020qy, Lambert20211b, Economides2021, Browning2022, Murphy2025}.
Other works fit a theoretical model to each experimental data point individually and obtain the cell property distribution afterwards \cite{Saadat2020, Adlung2021, Heyn2025, Klimek2025}.
Both approaches require a large number of model evaluations and so are restricted to mathematically simple models which can be evaluated within seconds on standard computers.

The recent progress in high-throughput microfluidic phenotyping techniques has provided large microscopy data sets from which mechanical properties of biological cells could, in principle, be infered with high accuracy.
In these cytometry devices, cells in suspension are pushed through microchannels or small slits and an image of the deformed cell is recorded \cite{Lange2015, Urbanska2018, Saadat2020, Kubankova2021v18, Recktenwald20226fr, Gerum2022, Walther2022, Wittwer2023, Williams2023, Zhou2025}.
From the latter automated image analysis procedures extract geometrical identifiers which can be correlated to the state of health of the cell \cite{Saadat2020, Gerum2022, Walther2022} and/or the donating patient \cite{Kubankova2021v18, Recktenwald20226fr, Williams2023}.
If available, a physics-based computational model can be matched with the data to quantify mechanical properties of the cell \cite{Saadat2020, Gerum2022, Wittwer2023}.
Although the development of mechanical cell models has progressed considerably, the simulation of a single experiment usually still requires hours even on state-of-the-art supercomputers.
This high computational cost limits comparisons between model and experiment to a single or very few cells and precludes the application of Monte-Carlo based methods to infer population-wide distributions from microfluidic measurements. 

Here, we introduce the likelihood functional as a novel statistical tool to infer cell property distributions from experimental data giving a means to quantify cell-to-cell variability in biological systems.
Importantly, by avoiding Monte-Carlo sampling, our approach is applicable even for models which are computationally expensive and which depend non-linearly on the model parameters.
It is furthermore designed specifically to deal with snapshot experimental data, where only a single image is taken for each cell and time-dependent trajectories are not available.

\section{Methods}
\subsection{Likelihood functional to infer cell property distributions}
To describe our novel methodology, we consider a generic experimental setup which allows one to measure a set of $N$ different observables $f_i$.
Collecting the $f_i$ into a vector $ \vec{ f}$ and carrying out measurements on a statistically significant number $N_ \mathrm{exp}$ of cells, a multidimensional histogram can be created which yields a multimodal distribution of observables $p _\mathrm{ exp } \left( \vec{ f} \right)$.
The measurement uncertainty is taken as normally distributed and independent for each observable with a width $ \sigma_i$ and must be specified based on the experimental apparatus.
In the scenarios we consider, however, the broadness of $p _\mathrm{ exp }$ is primarily caused by the inherent cell-to-cell variability and not by measurement error. 
We consider further a theoretical or computational model $\mathcal{M}$ depending on a set of $M$ parameters (i.e.\ cell properties and initial conditions) $q_j$. 
Collecting these again into a vector, the model mathematically corresponds to a mapping $\mathcal{M}: \vec{ q} \to \vec{ f}$ from parameter space to observable space.
In this approach, the cell-to-cell variability is reflected in a multimodal distribution $ \pi \left( \vec{ q} \right)$ of cell properties which we seek to infer from the measured observable distribution $p _\mathrm{ exp }\left( \vec{ f} \right)$. 

For this, we introduce the likelihood functional $\mathcal{L} \left( \pi \left( \vec{ q} \right) \right)$ as the key novel quantity that our method hinges on.
The likelihood functional integrates the classical likelihood over parameter space weighting each $ \vec{ q}$ with its probability of occurence given by $ \pi \left( \vec{ q} \right)$.
Thus, for a given $ \pi ( \vec{ q})$:
\begin{align}
	\mathcal{L} \left( \pi \left( \vec{ q} \right) \right) &= \prod_{ k =1} ^{ N _\mathrm{ exp } } \prod_{ i=1 }^N 
	\frac{ 1 } { \sqrt{ 2 \pi \sigma _i  ^2 } }
	\int d \vec{ q} \, \pi \left( \vec{ q} \right)  \mathrm{e}^{ - \frac{ \left( f_i ^{ (k) } -  \mathcal{M} \left( \vec{ q} \right)_i \right) ^2 } { \sigma _i ^2  } }. \label{likelihoodFunctional}
\end{align}
where $k$ enumerates the experimental snapshots and $i$ the observables.
Out of all possible property distributions, the one which maximizes $ \mathcal{L}$ is the one which best describes physical reality and will be termed $ \pi ^* \left( \vec{ q} \right)$.

To find $ \pi ^* \left( \vec{ q} \right)$ in practice, we first discretize the integral by a multidimensional trapezoidal rule.
The sampling points for the integration are given by a set of model evaluations over a discrete grid in parameter space. 
Next, the form of the distribution for each parameter is postulated based on physical knowledge.
For cell properties, typically a log-normal distribution with two characteristics, namely the mean $ \overline{q}_j$ and the width (standard deviation) $ s _j$, is assumed \cite{Furusawa2005}.
We here assume property distributions to be uncorrelated, but including correlations, if necessary, is straightforward.
Other parameter distributions such as, e.g., the initial position of the cell before entering the measurement channel can be estimated from knowledge of the experimental situation.
With this, eq.~[\ref{likelihoodFunctional}] can be evaluated and standard non-linear optimization algorithms can be used to determine the optimal characteristics and thereby the sought parameter distribution $\pi^*\left(\vec{ q} \right)$.

A key advantage of this approach is that no expensive model evaluations are required during the optimization of $\pi( \vec{ q})$ since the exponential factor in eq.~[\ref{likelihoodFunctional}], which contains the actual model evaluations $\mathcal{M}$, can be precomputed at the sampling points before the start of the optimization procedure.
As an important a-posteriori validation, the observable distribution can be predicted as
\begin{align}
	p _\mathrm{ pred }( \vec{ f}) = \prod_{ i=1 }^N \frac{ 1 } { \sqrt{ 2 \pi \sigma _i ^2  } } \int d \vec{ q} \, \pi^* ( \vec{ q}) \,  \mathrm{e}^{ - \frac{ \left( f_i   -  \mathcal{M} \left( \vec{ q} \right)_i \right) ^2 } { \sigma _i ^2  } } 		\label{eqn:prediction}
\end{align}
and be compared to the experimental data $p _\mathrm{ exp }( \vec{ f})$.


\subsection{Flow field calculations combining Lattice-Boltzmann and Boundary-Integral simulations}

Red blood cells flowing through microfluidic channels are highly sensitive to the initial shape with which they enter the measurement channel \cite{Noguchi_2010_PRE, Guckenberger_2018_phase, Agarwal2020}.
Such sensitivity also appears in the present RT-DC setup.
In contrast to earlier works \cite{Wittwer2023, Wang2023} who considered only the measurement channel and a short entry/exit section, we therefore choose to simulate the entire trajectory of an RBC from the reservoir to the exit of the measurement channel.

\begin{figure*}[h]
\begin{center}
\includegraphics[width=0.7\textwidth]{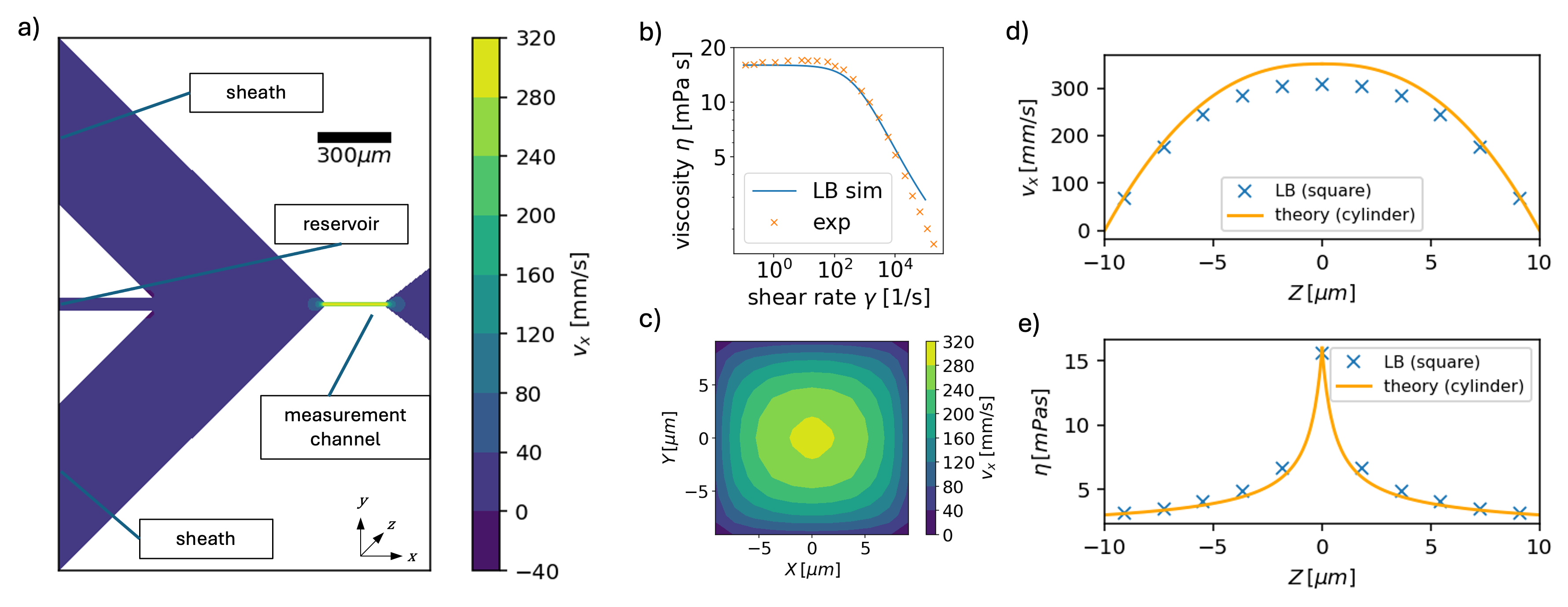}
\caption{(a) Flow field obtained by shear-thinning LB simulations for the entire RT-DC setup.
(b) Shear-thinning behavior of the methyl-cellulose carrier fluid in simulations and experiment \cite{Buyukurganci2023ci3} adjusted to a zero-shear viscosity of $\qty{15}{\milli\pascal\second}$.
(c) 2D cross-section of the flow field at the center of the measurement channel.
(d)+(e) velocity and viscosity profiles along the $z$-axis at the center of the measurement channel.}
\label{fig:checkChannelFlow}
\end{center}
\end{figure*}

The large size of this system, sketched in figure~\ref{fig:checkChannelFlow}~(a), poses a challenge to numerical simulations as the resolution needs to be coarse enough to fit the entire system and at the same time fine enough to resolve detailed changes in RBC shape.
We tackle this challenge by a two-step approach in which we first run a coarse Lattice-Boltzmann (LB \cite{Kruger_book}) simulation followed by a finely resolved boundary-integral (BI \cite{Pozrikidis_book}) simulation.

In the first step, we use LB to simulate the flow field through the entire system in the absence of the RBC as shown in figure~\ref{fig:checkChannelFlow}~(a).
As simulation tool we use our in-house code FluidX3D \cite{Lehmann20229da, Kellnberger2025, Kellnberger2026}.
The employed resolution is $\qty{0.95}{\micro\meter}$ per LB grid point.
Pressure boundaries are used at the three inlets and the outlet with pressures manually adjusted to obtain flow rates of $\qty{0.045}{\micro\liter\per\second}$ and $\qty{0.015}{\micro\liter\per\second}$ at the sheath and reservoir inlets, respectively, which directly match the experimental settings \cite{Kubankova2021v18}.
Bounce-back conditions are used to model rigid walls.
Simulations converged to a steady state after approximately 50,000 time steps.

The fluid used in the experiments was a 0.6\% methyl-cellulose buffer solution which is known to exhibit shear-thinning properties.
We capture shear-thinning in our LB simulations using a recently developed viscosity-shuffling procedure \cite{Kellnberger2025} following a Carreau-Yasuda relation \cite{Muller:2020co}:
\begin{align}
	\eta \left( \gamma \right) = \eta _\infty + \frac{ \eta _0 - \eta _ \infty } { \left( 1 + \left( K \, \gamma \right) ^{ a_1 } \right) ^{ \frac{ a_2 } { a_1 } } } 
\end{align}
with $K = \qty{2.1}{\milli\second}, a_1 = 0.96, a_2 = 0.387, \eta _ \infty = \qty{1}{\milli\pascal\second}, \eta _0 = \qty{15}{\milli\pascal\second}$.
The zero-shear viscosity of $\qty{15}{\milli\pascal\second}$ was chosen to match with the diluted solution in \cite{Toepfner2018}.
As shown in figure~\ref{fig:checkChannelFlow}~(b) these parameters reproduce well the experimental rheology measured in \cite{Buyukurganci2023ci3} with the zero-shear viscosity adjusted to $\qty{15}{\milli\pascal\second}$.
Elastic properties of the carrier fluid are neglected in our simulations.

As an illustration, figure~\ref{fig:checkChannelFlow}~(c) shows the flow field in the $y$-$z$ plane extracted at $x=\qty{280}{\micro\meter}$ after the start of the measurement channel. 
This is where we also compute the RBC deformation as detailed in section~\ref*{sec:contourComputation}.
Unfortunately, an analytical solution for a Carreau-Yassuda fluid in a square channel does not exist.
To provide an independent validation nevertheless, we use our recently developed semi-analytical theory for cylindrical channels \cite{Muller:2020co}.
In figure~\ref{fig:checkChannelFlow}~(d) and (e) we show that both the velocity and the local viscosity are in fair agreement with our LB simulations despite the different geometries.

In the second step, we simulate the trajectory and deformation of the RBC with the BI method as illustrated in figure~\ref*{fig:rbcTrajectory}.
BI methods solve the Stokes equation by discretizing only the fluid boundaries and thus do not require a volume grid even for simulating 3D domains \cite{Pozrikidis_book}.
In addition, they allow to easily incorporate an arbitrary background flow.
Here, we exploit these two properties by using the flow field computed in LB as a background flow for the RBC. 
As channel walls can in principle be omitted completely in this approach, we can use a fine resolution for the RBC shape (see also section \ref*{sec:simRBC}).
This simple approach, however, neglects hydrodynamic interactions between the RBC and the channel walls.
Due to the thinness of our system in $z$-direction, the RBC will ''feel'' only the top and bottom walls immediately above/below its current position.
To account for this, we include in our simulations two wall patches of size $40 \times 40 \si{ {\mu}m }$ located at $z=\pm \qty{10}{\micro\meter }$ just above and below the RBC.
As the RBC moves through the channel, the walls are shifted in $x$ and $y$ at every time step so that they follow the RBC position similar to approaches described in \cite{Roure2023}.
As the RBC enters the $20 \times 20 \si{ {\mu}m }$ measurement channel, it becomes strongly confined and interactions with all four walls are expected to become relevant.
Therefore, in this part, all channel walls are fully included in the BI simulation.
BI methods are \textit{a priori} not designed to handle shear-thinning fluids.
We therefore record from the LB simulations not only the flow field, but also the local viscosity at each point in the computational domain. 
In the BI simulations we use, at each time step, the local viscosity at the center of the RBC and thereby incorporate the shear-thinning effect also into our BI simulations.


\subsection{Forms of parameter distributions}
\label{sec:parameterForm}

Many cellular properties follow log-normal distributions \cite{Furusawa2005, Cai2013, Gerum2022, John2025}.
For the (dimensionless) cytosol and membrane viscosity we therefore assume the following form for the distribution
\begin{equation}
	\pi \left( q_j \right) = \frac{ 1 } { q_j \, w_j \, \sqrt{ 2 \pi  } }  \mathrm{e}^{ - \frac{ \left( \ln (q_j) - m _j \right) ^2 } { 2 w_j ^2  } }
\end{equation}
where the values $ m _j$ and $ w _j$ are related to the mean $\overline{q}_j$ and standard deviation $ s_j$ used in the main text as follows

\begin{align}
	w_j &= \sqrt{ \ln \left( \frac{ s_j ^2  } { \overline{q}_j ^2  } + 1 \right)}   \\	
	m_j &= \ln( \overline{q}_j ) - w_j ^2 /2.
\end{align}
Since our simulations show a very clear linear correlation between the starting position $y _\mathrm{ i }$ in the reservoir and the final position of the RBC in the measurement channel $y$, we choose the probability distribution for $y _\mathrm{ i }$ to follow precisely the experimentally observed distribution as shown in figure~\ref{fig:yCorr}.
The probability distribution of $z _\mathrm{ i }$ is flat.


\section{Results}
\subsection{Real-Time Deformability Cytometry for Red Blood Cells}

\begin{figure*}[h]
\includegraphics[width=\textwidth]{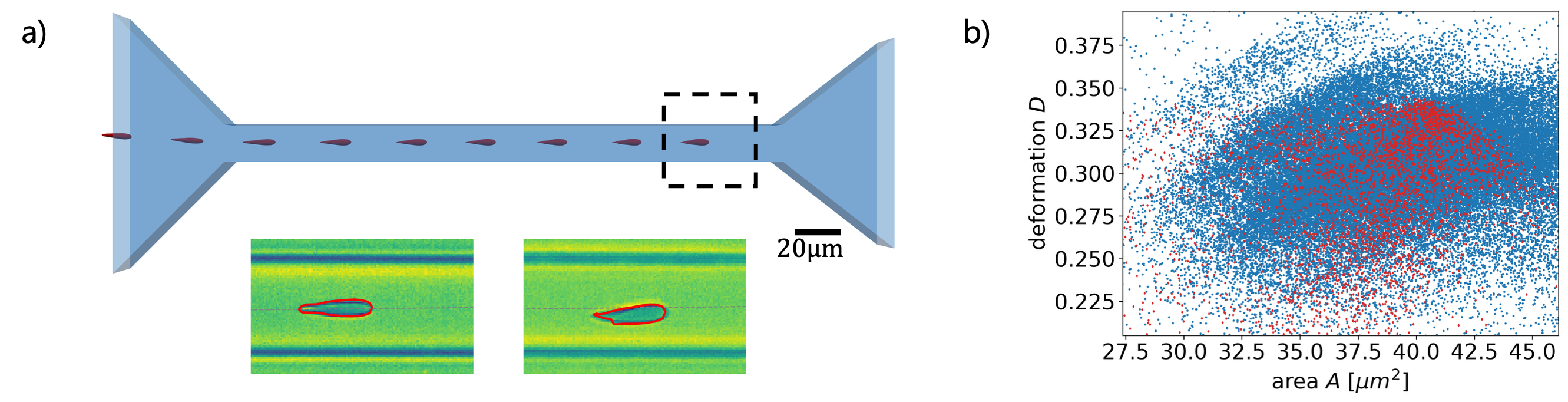}
\caption{Illustration of the RT-DC experiment \& simulations: (a) a diluted suspension of RBCs is pumped through the measurement channel; bottom images show the agreement between experiment \cite{Kubankova2021v18} and simulated images (red line) for two typical cell shapes.
(b) experimental data points (blue) from \cite{Kubankova2021v18} show appreciable scatter in the area-deformation plane;  simulations (red) with different model parameters (red dots) cover most of the relevant area.  }
\label{fig:RTDCmain}
\end{figure*}

To demonstrate the applicability of our approach to a real system, we choose Real-Time-Deformability-Cytometry (RT-DC) experiments \cite{Kubankova2021v18} which we use to determine distributions for the cytosol and membrane viscosity of red blood cells (RBCs).
In RT-DC, RBCs are pushed at high speeds through a 20x20 $ \si{ {\mu}m }$ measurement channel as illustrated in figure~\ref{fig:RTDCmain}~(a).
For each cell, a single image is recorded at the end of the measurement channel and the cell contour is automatically detected. 
Applying proper criteria, RBCs can be isolated from other blood cells such as leukocytes \cite{Kubankova2021v18}.
Two typical RBC shapes shown in figure~\ref{fig:RTDCmain}~(b) are an elongated shape with up-down symmetry flowing in the channel center and a slightly bent shape flowing above or below the center line. 
All experimental data in this contribution are taken from \cite{Kubankova2021v18} as detailed in section~\ref*{sec:DataCuration} of the Supplemental Material.

To create a digital twin of these RT-DC experiments, we conduct numerical simulations by a combination of Lattice-Boltzmann and boundary-integral methods taking into account shear-thinning properties of the carrier fluid as well as the strong dependence of RBC dynamics on initial conditions \cite{Noguchi_2010_PRE, Sinha_2015, Guckenberger_2018_phase, Agarwal2020} as detailed in Materials \& Methods below.
The mechanical model of the RBC membrane is by now fairly well established using the Skalak and Helfrich laws for shear and bending elasticity, respectively \cite{Gompper2016, Fedosov_Gekle_book}.
The shear and bending moduli can be determined with reasonable accuracy from static experiments such as micropipette aspiration as detailed in section~\ref*{sec:simRBC} of the Supplemental Material.
As we show in figure~\ref*{fig:varyKappa}, RT-DC experiments on RBCs are rather insensitive to these moduli and they are therefore considered as fixed values in the present context.

Both experimental shapes are well reproduced by our simulations (red lines in figure~\ref{fig:RTDCmain}~(a)).
From the contour, the enclosed area $A$ and a deformation parameter $D$ are computed as described in \cite{Kubankova2021v18}.
Figure~\ref{fig:RTDCmain}~(b) shows a histogram of about 65,000 experimental data points from 26 healthy donors taken from the dataset accompanying \cite{Kubankova2021v18} (see section~\ref*{sec:DataCuration}) in the area-deformation plane.
While the mean area and deformation are $\overline{A} =38.8 \si{{\mu}m}$ and $\overline{D}=0.304$, respectively, it is clear from figure~\ref{fig:RTDCmain}~(b) that there is an appreciable width in the distribution around these mean values.
In RBCs, this heterogeneity arises partly due to changes in the mechanical cell properties during aging.
In terms of the likelihood functional, we thus define the two-dimensional observable vector $ \vec{ f} = (A, D)$. 

\subsection{Cell-to-cell variability of red blood cell viscosities}

In contrast to the elastic shear and bending moduli, the dissipative, viscous properties of RBCs cannot be measured in static experiments.
Accordingly, quantitative values are known with significantly lower precision.
In a RBC, dissipation takes place in the inner cytosol as well as in the membrane itself and is therefore quantified by a cytosol viscosity $\mu _\mathrm{ cyto }$ and a membrane viscosity $\mu _\mathrm{mem }$ \cite{Fedosov_Gekle_book}.
The aim of this work is to use the likelihood functional to determine population-wide distributions of these two important cell properties.
We work in dimensionless values defining the viscosity contrast $ \lambda = \frac{ \mu _\mathrm{ cyto } } { \mu _\mathrm{ plasma } }$ and the Boussinesq number $ Bq = \frac{ \mu _\mathrm{mem } } { R \, \mu _\mathrm{ plasma } }$ as the dimensionless cytosol and membrane viscosity, respectively, normalized with $ \mu _\mathrm{ plasma } = \qty{1.2}{\milli\pascal\second}$ and a typical radius $R=\qty{4}{\micro\meter}$.

A complication for the simulations is that the initial position when the RBC leaves the reservoir, though measurable in principle, is not recorded in typical RT-DC experiments.
We therefore consider the initial positions $y _\mathrm{ i }$ and $ z _\mathrm{ i }$ as illustrated in figure~\ref*{fig:rbcTrajectory} as additional model parameters.

The parameter vector $ \vec{ q}$ in the likelihood functional is thus four-dimensional with $\vec{ q} = ( \lambda, Bq, y _\mathrm{ i }, z _\mathrm{ i })$.
For the first two properties we assume uncorrelated log-normal distributions with means $ \overline{ \lambda }, \overline{Bq}$ and widths $ s _ \lambda , s _\mathrm{ Bq }$ as detailed in Materials \& Methods below.
The distribution of $z _\mathrm{ i }$ is assumed to be homogeneous while the distribution of $y _\mathrm{ i }$ follows directly the observed distribution of $y$ positions in the measurement channel (see figure~\ref{fig:yCorr} in Supplemental Material).

\begin{figure}[h]
\begin{center}
\includegraphics[width=0.5\textwidth]{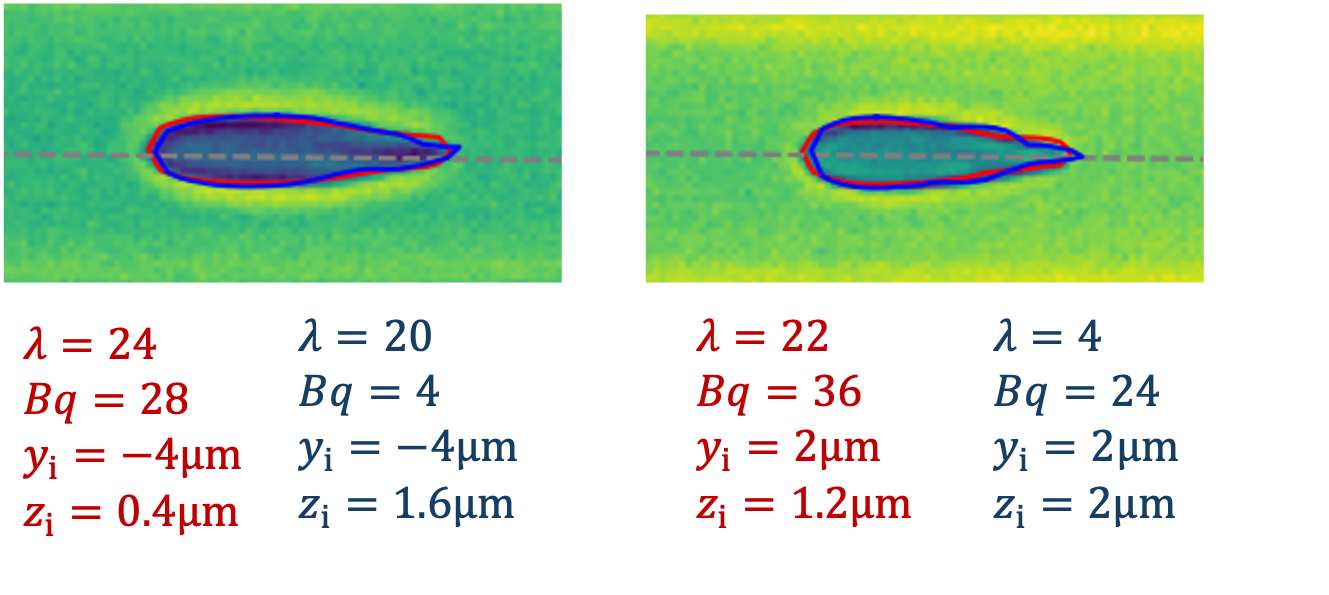}
\caption{The inverse mapping from a given experimental image to simulation parameters is not unique due to the high non-linearity of the mapping $\mathcal{M}$: a given experimental image can be well reproduced by two simulations with very different model parameters. 
This reiterates the need for a global inference approach such as our likelihood functional.}
\label{fig:nonlinear}
\end{center}
\end{figure}
We then conduct a series of in total 4,800 simulations spanning a grid of physically reasonable model parameters given in table~\ref{tab:simGrid} in Materials \& Methods.
From this, we obtain a cloud of simulation data points shown as red dots in figure~\ref{fig:RTDCmain}~(b) which covers the relevant experimental observable distribution.
The complexity of the RT-DC setup leads to a highly non-linear relation between the model parameters and the resulting RBC shape as illustrated in figure~\ref*{fig:nonlinear}.
This precludes parameter inversion approaches which rely on a direct one-to-one correspondence between observables and model parameters and demonstrates the need for a global statistical method.

\begin{table}
\begin{center}
\small
\begin{tabular}{|l|l|}
\hline
parameter &  \\
\hline
cytosol viscosity ratio $ \lambda $& 2, 4, 6, 8, 10, 12, 14, 16, 18, 20, 22, 24 \\
\hline       
membrane viscosity ratio $Bq$ & 4, 8, 12, 16, 20, 24, 28, 32, 36, 40 \\
\hline
$y _\mathrm{ i }$ [$\mu$m] & 0, 2, 4, 8, 12, 16, 20, 24\\
\hline
$z _\mathrm{ i }$ [$\mu$m] & 0.4, 0.8, 1.2, 1.6, 2.0 \\
\hline
\end{tabular}
\caption{The simulation grid used to cover the 4D parameter space.}
\label{tab:simGrid}
\end{center}
\end{table}

\begin{figure*}[h]
\includegraphics[width=\textwidth]{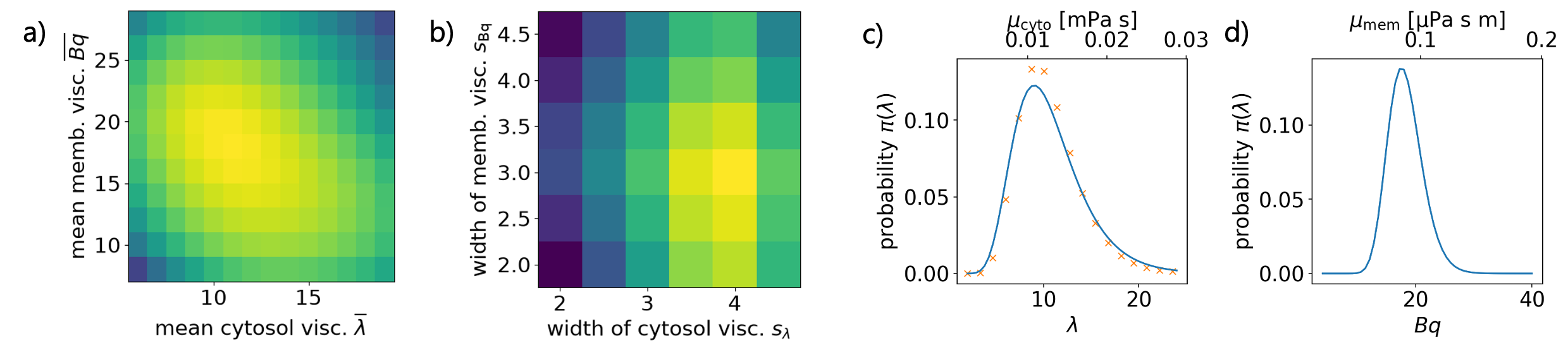}
\caption{The likelihood functional is able to obtain well-defined maxima for parameter distributions.
(a) fixing the widths $ s _ \lambda = 4$ and $s _\mathrm{ Bq } = 3$, a clear maximum for the means of the cytosol and membrane viscosity distributions is found. 
(b) fixing the means to $\overline{ \lambda } = 11, \overline{Bq}=18$, a similary clear maximum is found for the widths of the viscosity distributions.
Colors indicate $\log \left(  \mathcal{L} \right)$ in arbitrary units.
(c+d) After optimization of the likelihood functional, two log-normal distributions for cytosol and membrane viscosity red blood cells are predicted.
Their characteristics are $\overline{\lambda} = 10.7, \overline{Bq}=18.3, s_\lambda = 3.8, s _\mathrm{ Bq }=3.0$.
For the cytosol viscosity, recent direct measurements from \cite{John2025} are shown by the crosses and are in excellent agreement with our predictions. }
\label{fig:likeGridsDist}
\end{figure*}

Before embarking on a systematic maximization of the likelihood functional in eq.~[\ref{likelihoodFunctional}], we convince ourselves that the likelihood landscape indeed exhibits a clear and well-identifiable maximum.
For this, we first fix the widths $ s _ \lambda , s _\mathrm{ Bq }$ of the viscosity distributions and show in figure~\ref{fig:likeGridsDist}~(a) the likelihood when varying only the means.
A clear maximum in the region around $ \overline{ \lambda }  \approx 11$, $\overline{Bq}  \approx 18$ is visible.
Similarly, we show in figure~\ref{fig:likeGridsDist}~(b) that the likelihood functional exhibits a maximum not only for the means but also for the widths of the sought distributions.

%

Using these estimated maxima as starting guesses, we conduct Powell optimization on the distribution characteristics $\overline{\lambda}, \overline{Bq}, s _\lambda, s _{ Bq }$ to obtain the best-fit parameter distribution $ \pi ^*( \vec{ q})$.
The resulting log-normal distributions for cytosol and membrane viscosity are shown in figure~\ref{fig:likeGridsDist}~(c) and (d).
For the cytosol viscosity, recent experiments have provided a median value of $ \mu _\mathrm{ cyto }=\qty{12.75}{\micro\pascal\second}$ with a fairly broad log-normal distribution reflecting cell-to-cell variability \cite{John2025} which is in very good agreement with our approach.
For the membrane viscosity, a recent overview \cite{Fischer_RBC_2022} collecting data from various sources provided values roughly in the range $ \mu _\mathrm{ mem } = \qty{0.1}{ \micro\pascal\second\meter} - \qty{1}{ \micro\pascal\second\meter}$ which is somewhat higher but still compatible with our predicted distribution.

\begin{figure}[h]
\includegraphics[width=0.8\columnwidth]{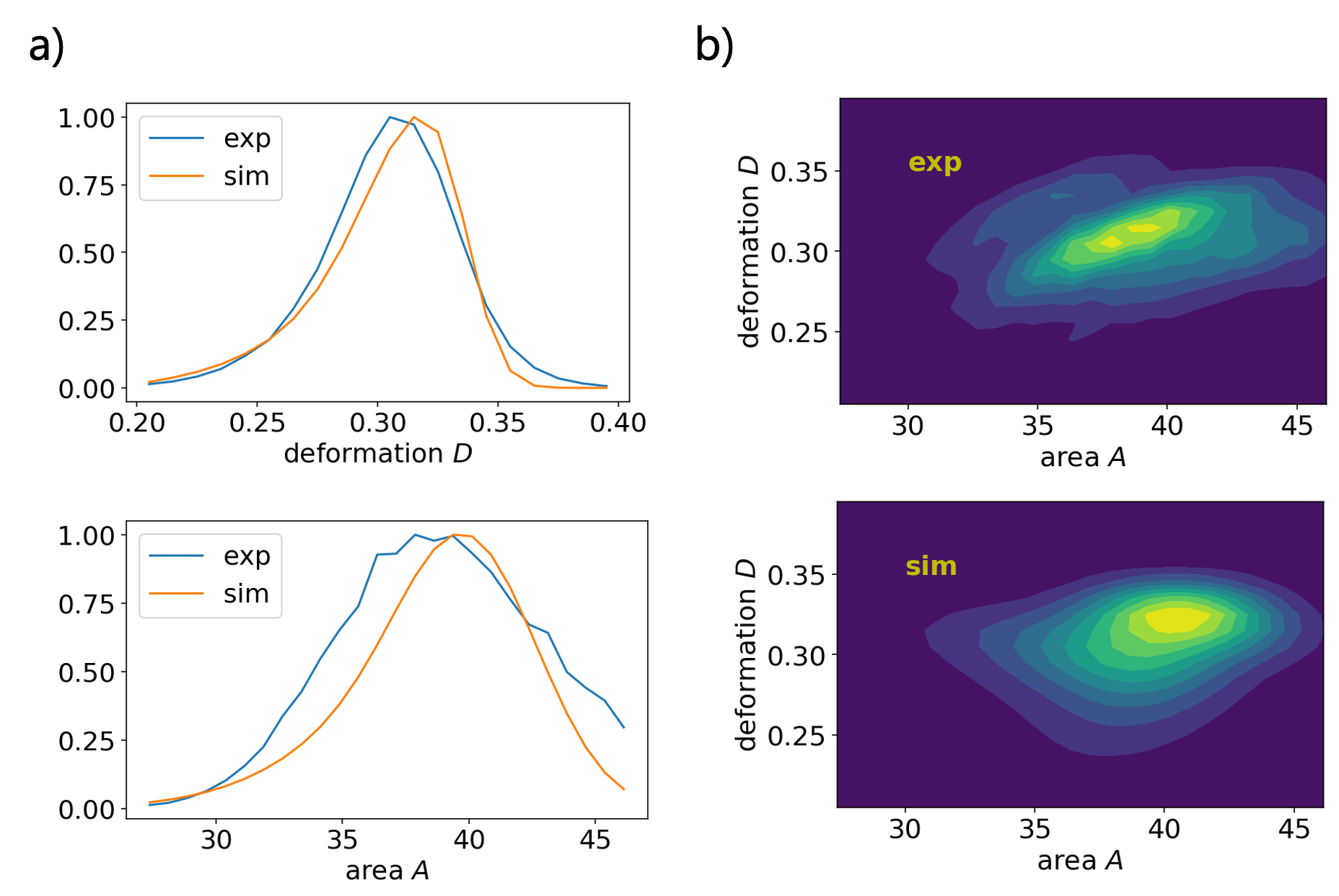}
\caption{
(a) Measured and predicted 1D distributions for area and deformation are in good agreement for the best-fit parameter distributions. 
(b) Comparison of measured and predicted 2D distributions in area-deformation space validates the likelihood functional prediction. }
\label{fig:optimal}
\end{figure}

An important advantage of the likelihood functional framework is that it allows for straightforward visual assessment of the quality of the inferred property distributions.
For this, we compute the predicted observable distribution $p _\mathrm{ pred }( \vec{ f})$ from eq.~[\ref{eqn:prediction}] and compare to the experimentally measured distribution $p _\mathrm{ exp }( \vec{ f})$. 
We start in figure~\ref{fig:optimal}~(a) with 1D histograms of area and deformation from the roughly 65,000 individual cell measurements \cite{Kubankova2021v18} and compare to the corresponding prediction from the likelihood functional.
For both observables we find convincing agreement.
An even stronger test for the likelihood functional approach is a comparison between measured and predicted observable distributions in the 2D area-deformation plane where we find again good agreement as shown in figure~\ref{fig:optimal}~(b).

\section{Conclusion}

We introduced the likelihood functional as a statistical tool to bridge the gap between high-throughput measurements and complex computational models.
This methodology allows to quantify biologically relevant cell-to-cell variability by inferring cell property distributions directly from measured observable distributions.
The key advantage of our method is that during the actual inference procedure no model re-evaluations are necessary which makes it applicable even for expensive computational models where Monte-Carlo estimation becomes unfeasible.
Using our method, we determined the distribution of red blood cell cytosol and membrane viscosities in a large population of healthy red blood cells. 
Deviations from these distributions may serve as an important indicator in medical diagnostics.

\section*{Acknowledgements}

This publication was funded by the University of Bayreuth Open Access Publishing Fund.
We acknowledge funding from Deutsche Forschungsgemeinschaft in the framework of FOR 2688 ''Instabilities, Bifurcations and Migration in Pulsating Flow'', Project B3 (349558021).
Calculations were performed using the festus-cluster of the Bayreuth Centre for High Performance Computing, funded by the Deutsche Forschungsgemeinschaft (DFG, German Research Foundation) - 523317330.

\newpage
\section*{references}
\bibliography{bibMod}

\newpage

\begin{center}
\Large Supplementary Material
\end{center}

\section{Data curation}
\label{sec:DataCuration}

Experimental data is taken from the dataset accompanying \cite{Kubankova2021v18} which is available from ZENODO under DOI {10.5281/zenodo.4737521}.
We focus on the dataset for healthy persons which contains data from 26 individuals. 
To estimate the distribution width of cytosol and membrane viscosity, we combine the data from all 26 patients into one dataset.
As the dataset contains all blood cells, we filter the raw data according to \cite{Kubankova2021v18} to select red blood cells: the cell area must lie between 15 and 50 $ \mathrm{{\mu}m}^2$, the aspect ratio between 2 and 10 and the area ratio between 1.0 and 1.08.
To match with simulation data, we further restrict the experimental data to cells whose area is between 27 and 47 $ {\mu}\mathrm{m}^2$ and whose $y$-position measured from the channel center is between -2 and 2 $ {\mu}\mathrm{m}$.

As observables we extract the columns ''area\_um'' as the cell area $A$ and ''deform'' as the cell deformation $D$ from the RT-DC raw data files.
We do not apply the correction proposed by \cite{Herold2017}  as we found that it does not notably influence our results.
To estimate the measurement error, we extract the raw contour of an image (available in the dataset) and perturb the $y$-position of each pixel randomly with a Gaussian noise of width 1 pixel.
Then we compute $A$ and $D$ for the perturbed image.
We repeat this procedure 100 times for the same image and take the standard deviations of all resulting $A$ and $D$.
Repeating this procedure for 100 randomly selected images and averaging over the standard deviations yields our estimate for the mesaurement error: $ \sigma _\mathrm{A} = \qty{2} {\micro\meter\squared}$ and $ \sigma _\mathrm{D} = 0.0114$.


\section{Red blood cell model}
\label{sec:simRBC}

\begin{figure}[h]
\begin{center}
\includegraphics[width=0.7\textwidth]{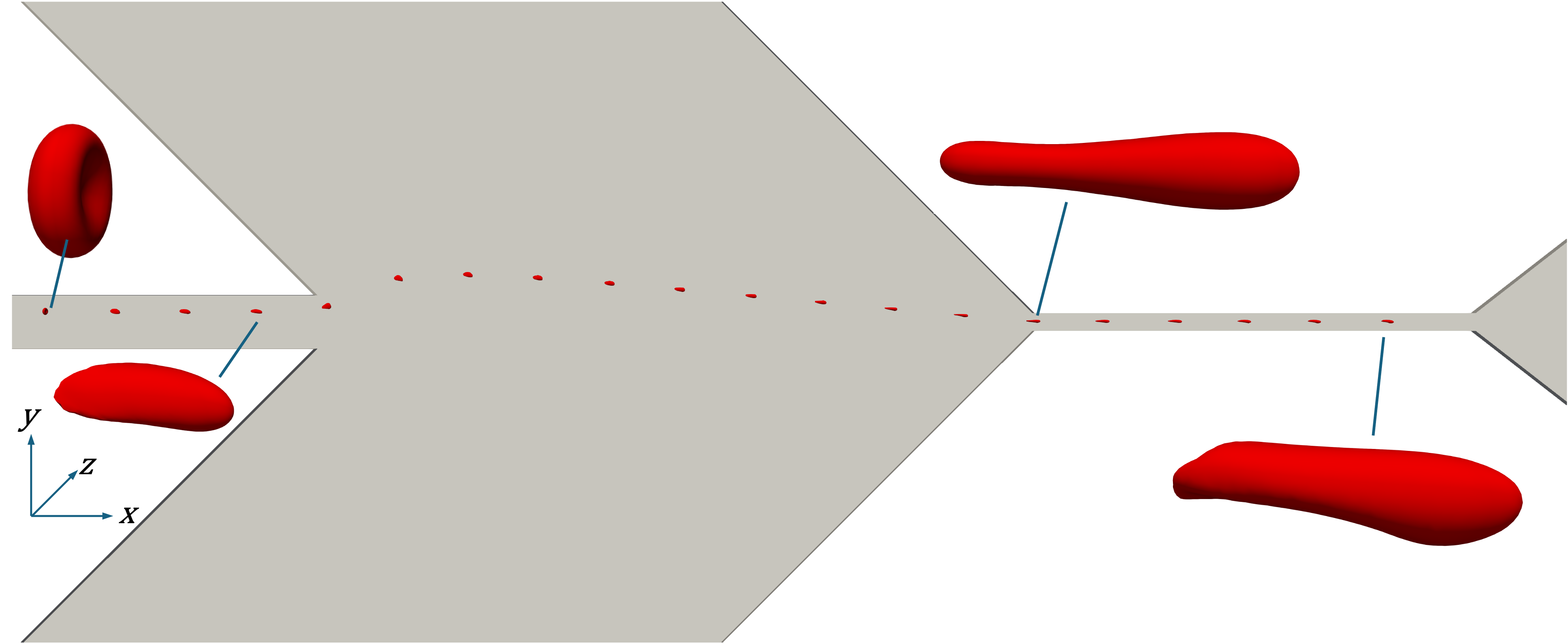}
\end{center}
\caption{A sample RBC trajectory as it passes through the RT-DC system. 
The RBC is simulated with a BI method using the flow field created by LB as a background flow.}
\label{fig:rbcTrajectory}
\end{figure}

RBC membrane elasticity is modeled using the Skalak and Helfrich laws for shear, area dilatation and bending resistance \cite{Fedosov_Gekle_book}.
The RBC membrane is triangulated and elastic forces are computed as described in our earlier publications \cite{Recktenwald2021, Guckenberger_2018_phase}.
The used moduli are $ \kappa _\mathrm{ s } = \qty{5e-6}{\newton\per\meter}$ for shear and $ \kappa _\mathrm{ b } = \qty{4e-19} {\newton\meter}$ with a reference curvature of $\qty{0.71}{ \per\micro\meter }$. 
The reference shape for shear elasticity is an oblate spheroid with the same surface area and one axis being 0.9 times shorter than the other two.

\begin{figure}[h]
\begin{center}
\includegraphics[width=0.9\textwidth]{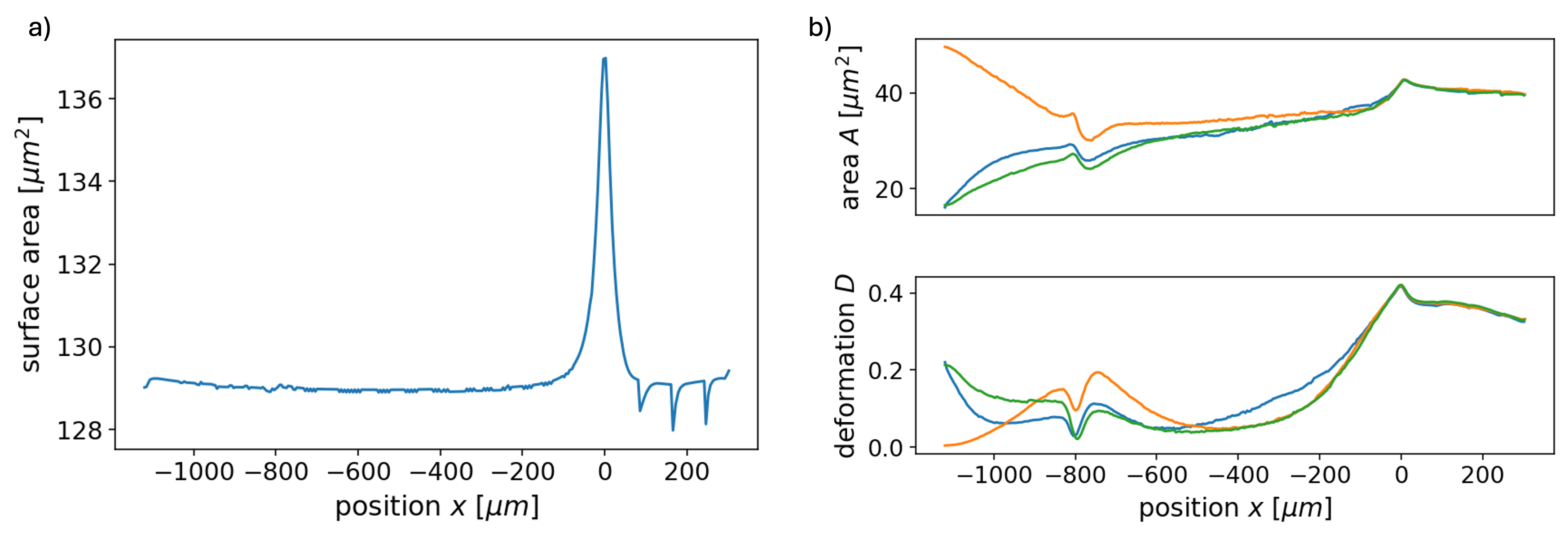}
\caption{(a) Variation of surface area during a typical simulation ($ \lambda =10, Bq=16, z _\mathrm{ init }=\qty{1.2}{\micro\meter}, y _\mathrm{ init } = \qty{8}{\micro\meter}$) does not exceed 6\%. 
(b+c) Variation of the starting orientation for the same simulation. 
Compared to the standard situation (''head on''; blue), we here show ''wheel'' with the symmetry axis of the initial discocyte parallel to the $z$-direction (orange) and ''flat'' with the symmetry axis parallel to $y$-direction (green).
}
\label{fig:robustnessSingle1}
\end{center}
\end{figure}

To ensure inextensibility of the RBC membrane, we use local (per triangle) as well as global penalty forces.
The forces are derived from a harmonic energy functional with moduli chosen large enough to ensure surface conservation and small enough to not decrease the time step too much.
Moduli are chosen lower in the beginning of the simulation and are increased in the region where the RBC enters the measurement channel and high stretching occurs.
The area conservation modulus in the Skalak law is set to zero.
Figure~\ref{fig:robustnessSingle1}~(a) shows that the change in surface area during a simulation is typically around 6\%.

Volume conservation is ensured strictly by rescaling the RBC volume at every time step.
Membrane shear viscosity is included by extending the approach of \cite{Diaz2001} to three dimensions as in \cite{Mayr2026}.
To ensure a correct computation of the velocities from the BIM equations, we include the viscous force contributions on the left-hand side of the BIM equation.
Membrane dilatational viscosity is set equal to membrane shear viscosity $ \mu _\mathrm{ mem }$, but we found that its influence on the results is insignificant.

The initial shape follows the discocyte functional form given by \cite{Evans_1972} in their equation (7) with the slightly modified parameters $R_0=3.97 \si{ {\mu}m }, C_0=0.81 \si{ {\mu}m }, C_2=6.97 \si{ {\mu}m }, C_4=-4.41 \si{ {\mu}m}$. 
The corresponding surface area and volume are 129$\si{{\mu}m^2}$ and 85fl and are thus within the range of the limited experimental data available \cite{Evans_1972, Fung1981, Byun2012, Park2016, Namvar2021}.
We note that most likely the quiescent discocyte shape of RBCs also exhibits cell-to-cell variability \cite{Fung1981} but we restrict ourselves here to a single shape in order to make the parameter space not too large.

The initial orientation of the RBC is ''head on'' to the flow as shown in figure~\ref{fig:rbcTrajectory}.
In figure~\ref{fig:robustnessSingle1}~(b) we show that choosing different starting orientations does not influence the outcome of the simulations as the RBC quickly reorients and adapts to the flow.

\subsection*{Insensitivity to elastic moduli}
\label{sec:insensitivityElastic}

\begin{figure}[h]
\begin{center}
\includegraphics[width=\textwidth]{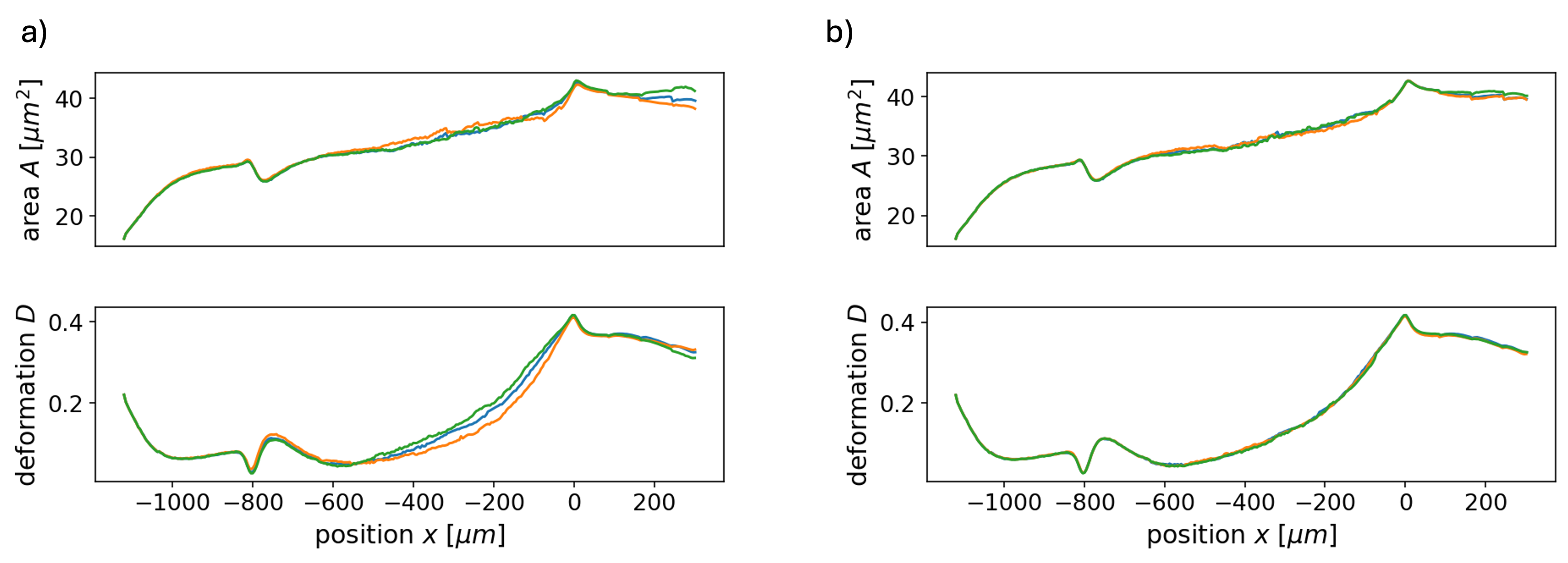}
\caption{Varying the elastic moduli for a typical simulation ($ \lambda =10, Bq=16, z _\mathrm{ init }=1.2\si{{\mu}m}, y _\mathrm{ init } = 8 \si{ {\mu }m }$) shows that the system is not notably influenced by elastic forces.
(a) Starting from the standard value of $ \kappa _\mathrm{ s } = 5 \cdot 10^{-6} N/m$ (blue) for the shear modulus, we increased/decreased $ \kappa _\mathrm{ s }$ to $7.5 \cdot 10^{-6} N/m$ (orange) and $3.75 \cdot 10^{-6} N/m$ (green), respectively, without any significant change.
(b) Increasing/decreasing the bending modulus in a similar way does not affect the results either.
}
\label{fig:varyKappa}
\end{center}
\end{figure}

We investigated whether the elastic moduli, $ \kappa _\mathrm{ s }$ for shear and $ \kappa _\mathrm{ b }$ for bending, may be required as additional components for the property vector $ \vec{ q}$ in the main text, in particular since earlier work \cite{Saadat2020} showed that the shear modulus in a RBC population does exhibit a non-negligible distribution.
We found, however, that our system is fairly insensitive to variations of the shear and bending moduli $ \kappa _\mathrm{ s }$ and $ \kappa _\mathrm{ b }$, respectively, as shown in figure~\ref{fig:varyKappa}.

\subsection*{Remeshing}

The initial mesh is created by starting from an icosahedron and subdividing the triangles \cite{Silva2024sr}.
After each subdivision step, nodes are moved radially outwards from the origin to create a smooth approximation of a sphere with radius $R$.
The initial mesh possesses 5120 triangles and 2562 nodes, each of which can be identified uniquely via its polar angle $ \theta $ and azimuthal angle $\varphi $ on the original sphere. 
We will retain this $ \theta , \varphi $ system as a surface parametrization throughout the simulation.
Next, the sphere is deformed into its initial discocyte as well as its spheroidal reference shape.

\begin{figure}[h]
\begin{center}
\includegraphics[width=0.8\textwidth]{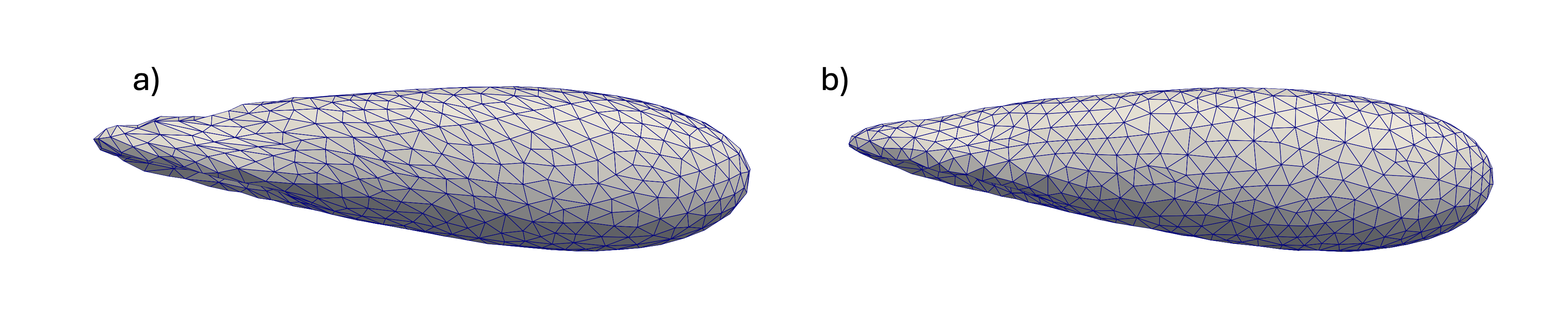}
\caption{Illustration for the improvement of mesh quality due to remeshing. Original mesh (a) and remeshed version (b).}
\label{fig:remeshIllu}
\end{center}
\end{figure}

\begin{figure}[h]
\begin{center}
\includegraphics[width=\textwidth]{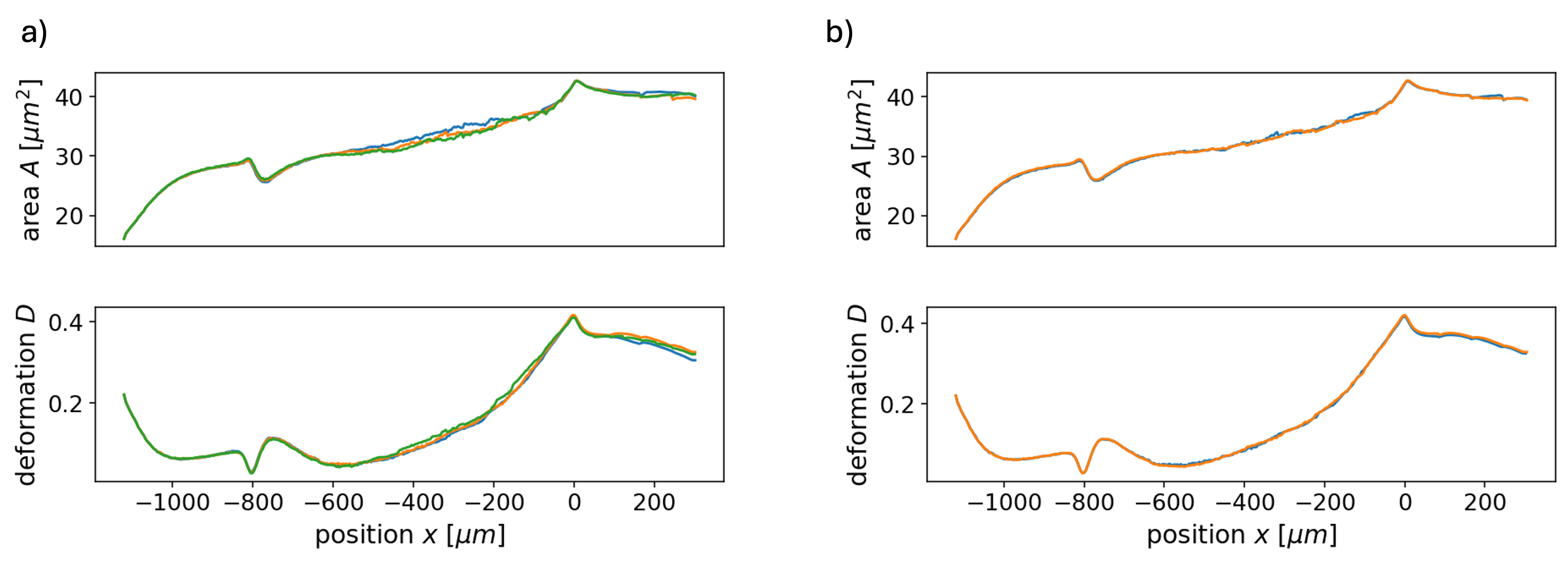}
\caption{
Evolution of the observables $A$ and $D$ along the RBC trajectory ($ \lambda =10, Bq=16, z _\mathrm{ init }=1.2\si{{\mu}m}, y _\mathrm{ init } = 8 \si{ {\mu }m }$) when changing the mesh properties.
(a) Varying the number of remeshing steps from 12 (blue) to 23 (orange) and finally to 39 (green).
(b) The standard (blue) resolution uses around 750 nodes and 1400 triangles, while the fine resolution (orange) uses around 1200 nodes and 2500 triangles. 
The values of the observables are not significantly influenced in either case.}
\label{fig:remesh}
\end{center}
\end{figure}

The long trajectory and the high velocities lead to significant deformation of the RBC during its trajectory.
To prevent mesh degradation and to keep the simulation stable, we remesh the cell surface about 10-15 times during its trajectory.
To maintain a proper link between the reference shape and the actual deformed RBC, remeshing is carried out using the $ \theta , \varphi $ parametrization introduced above.
A remeshing step then proceeds by first expanding the current as well as the reference shape into spherical harmonics using these $ \theta , \varphi $ as coordinates.
We then use the anisotropic BAMG remeshing algorithm implemented in gmsh \cite{Geuzaine2009} to construct a new mesh in $ \theta , \varphi $.
The node densities in each direction are chosen according to the local stretch tensor at each node.
Using the previously stored spherical harmonics coefficients, we can recover both the current and the corresponding reference shape for each triangle. 
This leads to a new surface discretization without (or with only very few) distorted triangles.
Figure~\ref{fig:remeshIllu} illustrates a remeshing step.
We checked the robustness by increasing/decreasing the number of remeshing steps and found that it did not affect notably our simulation results as shown in figure~\ref{fig:remesh}~(a).
Due to the remeshing, the discretization changes during the trajectory with typical values around 750 nodes and 1400 triangles.
In figure~\ref{fig:remesh}~(b) we show that a higher resolution does not affect the simulation outcome.

\section{Numerical integration}
\label{sec:numericalIntegration}

To evaluate the likelihood functional in eq.~[\ref{likelihoodFunctional}] of the main text, we discretize the integral in $ \vec{q}$-space and use nested trapezoidal integration. 
We checked that higher-order integration schemes such as Simpson do not notably affect the results.
The nodes for the integration in $ \vec{ q}$-space are chosen on the grid given in table~\ref*{tab:simGrid} of the main text.
Each node corresponds to one conducted simulation.
Integrating over parameter space distinguishes our method from other so-called surrogate approaches which interpolate, rather than integrate, in parameter space \cite{Economides2021}.
With this grid, we cover the range of values used in current literature for the cytosol \cite{Yoon2009, John2025} as well as for the membrane viscosities \cite{Guglietta2020, Matteoli2021, Gurbuz2023}.
More importantly, the resulting observables capture nearly the entire range of experimental observables as shown in figure~\ref*{fig:RTDCmain}~(b) of the main text.

\begin{figure}[h]
\begin{center}
\includegraphics[width=\columnwidth]{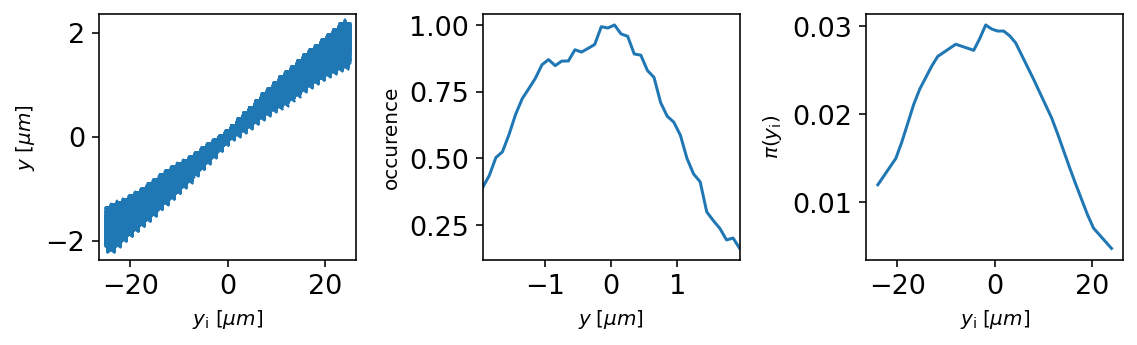}
\caption{(a) The simulations show a linear correlation between the starting position $y _\mathrm{ i }$ in the reservoir and the final position of the RBC in the measurement channel $y$.
(b) Histogram of the experimental $y$ in the measurement channel. Due to the sheath flow, cells are mostly centered.
(c) The probability distribution $ \pi ( y _\mathrm{ i })$ is fixed to follow the experimentally observed $y$-distribution.}
\label{fig:yCorr}
\end{center}
\end{figure}

Since the integrand in the likelihood functional is highly peaked around the maximum of the parameter distribution $ \pi\left(\vec{ q}\right)$, brute force integration over a roughly equidistant grid would lead to sizable numerical errors.
Before carrying out the integration, we therefore refine the node density around the maximum of the parameter distribution in $ \vec{ q}$-space.
The function values $\mathcal{M}\left(\vec{ q}\right)$ at these nodes are obtained by multidimensional linear interpolation from the neighboring simulations.
For the $ \lambda $ and the $ Bq$ direction, we use 40 integration points clustered around the maximum of the respective log-normal distributions.
For the $y _\mathrm{ i }$ direction, we use 40 equally distributed points since $ \pi ( y _\mathrm{ i })$ exhibits a broad maximum and no sharp peak as shown in figure~\ref*{fig:yCorr} of the main text.
Finally, for the $z _\mathrm{ i }$ direction, we directly use the simulated values without further interpolation.

Due to the outer product in the likelihood functional (equation (\ref{likelihoodFunctional}) of the main text), the integral needs to be evaluated $N _\mathrm{ exp }$ times.
Since in our case $N _\mathrm{ exp }\approx 65,000$, this is computationally not feasible.
We therefore bin the experimental data points into 20x20 bins in $ \vec{ f}$-space and perform the product over these bins with each bin weighted by the number of experimental data points contained.

\begin{figure}[h] \begin{center}
\includegraphics[width=0.7\textwidth]{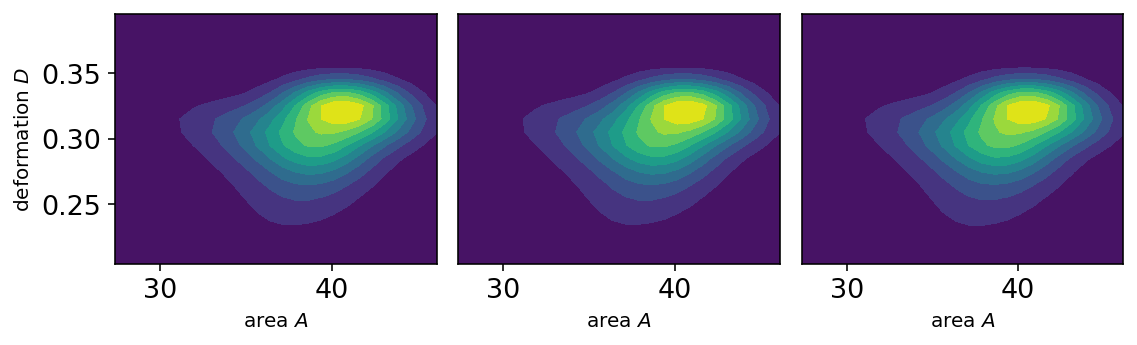}
\caption{Predicted observable distributions as in figure~\ref{fig:optimal}~(d) of the main text are robust when varying technical parameters during likelihood functional computation.
(a) Coarsening experimental data into 15x15 instead of 20x20 bins.
(b) Using 20 instead of 40 interpolation points for $y _\mathrm{ i }$.
(c) Using 30 instead of 40 interpolation points for the $ \lambda $ and $Bq$ intergration.}
\label{fig:finalPrediction_robustness}
\end{center} \end{figure}

We checked that neither the interpolation in $ \vec{ q}$-space nor the binning of the experimental data points in $ \vec{ f}$-space affected the predictions of the model as shown in figure~\ref{fig:finalPrediction_robustness}.

 \section{Contour and observable computation}
 \label{sec:contourComputation}
 
In RT-DC experiments, the observables area $A$ and deformation $D$ are computed from 2D microscopy images \cite{Urbanska2018, Kubankova2021v18}.
To obtain 2D projections from our 3D simulation data, we assume that the focal plane of the experimental images is in the channel center at $z=0$ and that the focal width is about 1${\mu}\mathrm{m}$.
We then cover the simulated RBC surface with a dense cloud of points by placing 20 random points within each surface triangle.
Out of this cloud, we select those points whose $z$-position is located within $\pm 1{\mu}m$ around $z=0$ and project them into the $x$-$y$-plane.
To properly compute the contour from the resulting 2D point cloud, we group all points according to their $x$ position into 25 boxes, equally spaced between the smallest and largest $x$ position.
In each box, the point with the largest (smallest) $y$ position is taken to represent the upper (lower) surface at the $x$ position of the box center.
This procedure robustly leads to an ordered set of points representing the discretized contour from which the area $A$ and the circumference $l$ can be computed and combined into the deformation parameter \cite{Kubankova2021v18}

\begin{align}
	D = \frac{ 1 - 2 \sqrt{ \pi \, A } } { l }.
\end{align}

\begin{figure}[h]
\begin{center}
\includegraphics[width=0.8\textwidth]{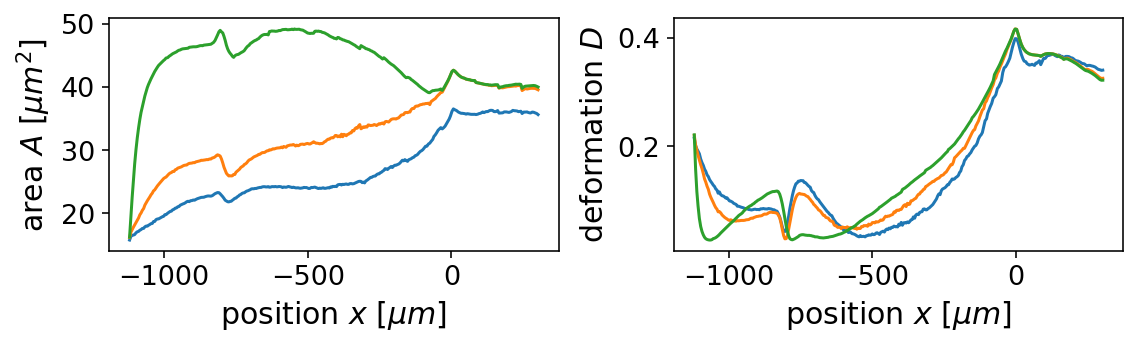}
\caption{Evolution of the observables $A$ and $D$ along the RBC trajectory when analyzed with different focal widths of 0.4${\mu}m$ (blue), 1${\mu}m$ (orange) and infinity (green). }
\label{fig:focalWidth}
\end{center}
\end{figure}

We checked that different choices for the number of random points per triangle as well as the number of boxes do not significantly influence the results.
Similarly, as shown in figure~\ref{fig:focalWidth} the focal width has only a minor influence on the calculation at the observation position unless it is chosen unrealistically small.

In the simulations, the full RBC trajectory is available while RT-DC experiments only record a single snapshot at the end of the measurement channel.
From each simulated RBC trajectory we therefore analyze the frame which is located 280 ${\mu}m$ behind the start of the measurement channel.


\newpage


\section{Correlation of initial and final position in $y$-direction}

\end{document}